\documentstyle[12pt]{article}

\def\thefootnote{\fnsymbol{footnote}}

\newcommand{\eq}{\begin{equation}}
\newcommand{\en}{\end{equation}}
\newcommand{\eqa}{\begin{eqnarray}}
\newcommand{\ena}{\end{eqnarray}}

\newcommand{\NP}[1]{Nucl.\ Phys.\ {\bf #1}}
\newcommand{\PL}[1]{Phys.\ Lett.\ {\bf #1}}

\newcommand{\PR}[1]{Phys.\ Rev.\ {\bf #1}}
\newcommand{\PRL}[1]{Phys.\ Rev.\ Lett.\ {\bf #1}}

\begin{document}

\hskip 11.5cm \vbox{\hbox{DFTT 8/94}\hbox{March 1994}}
\vskip 0.4cm
\centerline{\Large\bf The Spatial String Tension in High}
\centerline{\Large\bf Temperature Lattice Gauge Theories}
\vskip 1.3cm
\centerline{ M. Caselle and A. D'Adda
\footnote{email: CASELLE@TO.INFN.IT, DADDA@TO.INFN.IT}}
\vskip .6cm
\centerline{\sl  Dipartimento di Fisica
Teorica dell'Universit\`a di Torino}
\centerline{\sl Istituto Nazionale di Fisica Nucleare,Sezione di Torino}
\centerline{\sl via P.Giuria 1, I-10125 Turin,Italy}
\vskip 2.5cm

\begin{abstract}
We develop some techniques which allow an analytic evaluation of
space-like observables in high temperature lattice gauge theories.
We show that such variables are described extremely well by
dimensional reduction. In particular, by using results obtained in the
context  of ``Induced QCD'', we evaluate the contributions  to space-like
observables coming from the  Higgs sector of the dimensionally reduced
 action, we find that they are of higher order in the coupling constant
compared to those coming from the space-like action and hence neglegible
near the continuum limit.
In the case of SU(2) gauge theory our results agree with those obtained
through Montecarlo simulations both in (2+1) and (3+1) dimensions and
they also indicate a possible way of removing the gap between the two
values of $g^2(T)$ recently appeared in the  literature.
\end{abstract}
\noindent
\vfill
\eject

\newpage

\setcounter{footnote}{0}
\def\thefootnote{\arabic{footnote}}

\section{Introduction}

It is by now
widely accepted  that  (d+1) dimensional
Lattice Gauge Theories drastically simplifies at high temperature
and can be described by
some effective d-dimensional model. This approach, which is
usually called ``dimensional reduction'', is rather non-trivial.
For instance, it can be shown that the naive assumption of a complete
 decoupling of the ``non-static modes'' (namely, the degrees of
freedom, in the compactified ``temperature'' direction) does not hold in
general and that these non-static modes induce some well defined
interaction in the static sector.
The predictions of dimensional reduction  have been successfully
tested recently in the case of (3+1) dimensional SU(2)~\cite{lmr}
and SU(3)~\cite{klmpr1}  LGT, by looking at ``time-like'' observables
(correlations of Polyakov loops).

 At the same time there has been a
renewed interest, in the last year, in the study of space-like
Wilson loops at high
temperature, both in (2+1) and (3+1) dimensions and both for continuous
(SU(2) and SU(3))~\cite{t1,bfhks,klmpr2} and discrete ($Z_2$)
{}~\cite{cfggv} gauge groups. Even if
the space-like
string tension extracted from these Wilson loops has nothing to do with
the interquark potential (as a matter of fact, it
is different from zero, and actually
increases with the temperature, also in the deconfined region)
there are various
reasons of interest in its behaviour.
First, it can be used to predict various
physical features of the theory, like the temperature dependence of
quark-antiquark correlations in the space-like directions~\cite{koch}.
Second, it can be related to the thickness of the flux tube joining the
quark-antiquark pair at {\sl zero temperature}~\cite{t1,t2,cfggv} .
Third it can be used as a further independent check of the dimensional
reduction approach in the case of space-like observables.
This last point was addressed in~\cite{bfhks} (see
also~\cite{k-lat93} for a comprehensive review) where, by using a set of very
precise data on SU(2), an impressive agreement between the space-like
string tension in (3+1) dimensions and the zero temperature
string tension in 3 dimensions was found. The only missing  point for a
complete success of the dimensional reduction program was a misagreement
(roughly of a factor of two) between the coupling constants quoted
in~\cite{lmr} and that of~\cite{bfhks}.

In this paper we want to further pursue this analysis by using some
recent results obtained in a completely different
framework, namely the so called ``induced QCD''~\cite{KM}.
We shall show in
sect.2 and 3 that, within this framework, it is possible to define
directly on the lattice a dimensional reduction approach for high
temperature LGT's~\cite{CAP}. Moreover we shall show that, as far as
space-like observables are concerned, further simplifications occur,
part of the dimensionally reduced effective action can be neglected,
thus allowing in particular cases to make explicit, analytic
calculations. We shall then test our predictions
in the case of the SU(2) model in (2+1) (sect.4) and (3+1) (sect.5)
dimensions. In the latter case we shall give some argument to remove
the apparent discrepancy between the results of ref.~\cite{lmr} and
\cite{bfhks}.

\section{General Setting and Notations}

Let us consider a pure gauge theory with gauge group $SU(N)$ defined
on a $d+1$ dimensional cubic lattice. In order to describe a finite
temperature LGT we must take periodic boundary conditions in one
direction (which we shall call from now on ``time-like'' direction),
while the boundary conditions in the other $d$ direction (which we shall
call ``space-like'') can be chosen freely. Let us take a lattice of
 $N_t$  ($N_s$) spacings in the time (space) direction.
The theory will contain only gauge fields described by the link
variables $U_{ n;i} \in SU(N)$ where $ n \equiv (\vec x,t)$ denotes the
 space-time position of the link and $i$ its direction.
It is useful to choose
different couplings in the time and space directions. Let us call them
$\beta_t$ and $\beta_s$ respectively. Let us take the simplest choice
for the lattice gauge action, namely the Wilson action:

\eq
S_W=\sum_{n}~\frac{1}{N}~Re\left\{\beta_t\sum_i~{\rm Tr_f}(U_{n;0i})
+\beta_s\sum_{i<j}~{\rm Tr_f}(U_{n;ij})\right\}~~~,
\label{wilson}
\en
where ${\rm Tr_f}$ denotes the trace in the fundamental representation
and $U_{n;0i}$ ($U_{n;ij}$) are the time-like (space-like)
plaquette variables, defined as usual:

\eq
U_{n;ij}=U_{ n;i}U_{n+ \hat i;j}
U^\dagger_{n+\hat j;i}U^\dagger_{n;j}~~~.
\en

In the following we shall call $S_s$ ($S_t$) the space-like (time-like)
part of $S_W$.
$\beta_s$ and $\beta_t$ are related to the  (bare) coupling constant $g$
 and the
temperature $T$ of the gauge theory by the usual relations:

\eq
\frac{2N}{g^2}=a^{3-d}\sqrt{\beta_s\beta_t},
\hskip 1cm
T=\frac{1}{N_ta}\sqrt{\frac{\beta_t}{\beta_s}}
\label{couplings}
\en
where $a$ is the space-like lattice spacing while $\frac{1}{N_tT}$ is
 the time-like spacing. The two are related by the adimensional ratio
$\epsilon\equiv \frac{1}{N_tTa}$.
We can solve the above equations in terms of $\epsilon$ as follows:
\eq
\beta_t=\frac{2N}{g^2\epsilon}a^{d-3}
\label{betat}
\en
\eq
\beta_s=\frac{2N\epsilon}{g^2}a^{d-3}~~.
\label{betas}
\en

In a finite temperature discretization it is possible to define
gauge invariant variables which are topologically non-trivial loops,
closed due to the periodic boundary conditions in the time directions.
The simplest choice is the Polyakov loop defined as follows:

\eq
P(\vec x)= {\rm Tr} \prod_{t=1}^{N_t}(U_{\vec x,t;0})
\label{polya}
\en
where ${\vec x}$ labels the space coordinates  of the lattice sites.
Moreover, an important feature of the finite temperature theory with
respect to the zero temperature case is that it has a new global
symmetry (independent from the gauge symmetry) with symmetry group the
center $C$ of the gauge group (in our case $Z_N$). The Polyakov loop
turns out to be a natural order parameter for this symmetry.

For $d>1$ these theories admit a
deconfinement transition at $T=T_c$.
In the following we shall be interested in the high temperature phase
($T>T_c$).
It is possible to obtain, just from the definition itself of the model,
some general properties of this high temperature regime (see for
instance~\cite{sy}). In this region the symmetry with respect of the
center of the group is broken, the theory is deconfined, the Polyakov
loop has a non-zero expectation value and, what is more important, it is
an element of the center of the gauge group. In the following this $Z_N$
degeneracy will not play any role and we shall assume to have lifted it,
by selecting the vacuum in which
\noindent
$<P(\vec x)>={1}$.

The simplest approach to the description of the high temperature phase
is the so called ``{\sl complete} dimensional reduction''
according to which the degrees of freedom in the compactified
time direction (the non-static modes) decouple~\cite{ap}. The resulting
 theory is a $d$-dimensional gauge theory coupled to a scalar field in
the adjoint representation, and the corresponding
action is the sum of the purely space-like part $S_s$ and a new term
$S_h$ which is the remnant of the $S_t$ term:
\eq
S_h(m_0)=\frac{\beta_h}{N}~{\rm Tr_f}\sum_{\vec x}
\left(-m_0^2 \phi(\vec x)^2+
\sum_{i=1}^d U_{\vec x;i}\phi(\vec x)U^\dagger_{\vec x;i}
\phi(\vec x+\hat i)\right)
\label{higgs}
\en
where $\phi(x)$ is an Hermitian $N \times N$ matrix and
$m_0^2=d$.

 However, it is by
now clear that such a complete dimensional reduction does not take place
in general, that non-static modes cannot be neglected and that they
induce some well defined interaction terms in the static sector.
These induced interactions can be explicitly
evaluated and taken into account (see~\cite{lmr}, and references
therein).
They amount to a self-interaction term for the scalar field $\phi$:
\eq
S_{ns}= -\frac{\beta_h}{N}\sum_{\vec x} \left\{\frac{h}{2}~{\rm Tr_f}
\phi^2(\vec x) +
\frac{k_1}{2}~{\rm Tr_f} \phi^4(\vec x)\right\}~~~.
\en
The detailed derivation of this term and the values of $h$ and $k$
in the $N=2,3$, $d=3$ case, can be
found in ref.~\cite{lmr,klmpr1} (for $N>3$ a term of the type
$({\rm Tr_f} \phi^2(\vec x))^2$ should be also taken into account).

Notice that the quadratic term can be absorbed in a redefinition
of the mass term $m^2\equiv m_0^2+h/2=d+h/2$.

Following the literature on the
subject~\cite{lmr,k-lat93}, we shall simply call this procedure
``dimensional reduction'' and the resulting effective action:
$ S_s+S_h+S_{ns}$ the ``dimensionally reduced action''.

The $S_{h}$ and $S_{ns}$ contributions to the action are very important
if one studies time-like observables, such as the correlations of
Polyakov loops, but it turns out
that they are almost negligible if space-like observables are studied.
This has been anticipated in~\cite{k-lat93} and will be also shown
in the following.

\section{The Strategy}

Our starting point is the following assumption.

It is clear from equations (\ref{couplings},\ref{betat},\ref{betas})
that we have a
residual freedom in the choice of our lattice regularization, since
several choices of $(N_t,\beta_t,\beta_s)$ correspond to the same
set of $(g,T)$ values.
We assume that there is a region in the $(g^2,T)$ plane in which
we are allowed to fix  this residual freedom
 by choosing $N_t=1,~~
\beta_t=\frac{2NT}{g^2}a^{d-2},~~
\beta_s=\frac{2N}{g^2T}a^{d-4}~~.$

The resulting action is exactly of the type one would obtain within the
framework of a {\sl complete}
dimensional reduction (see ref.~\cite{CAP}). In fact $S_s$ is
not affected by the
transformation (except for the rescaling of $\beta_s$), while $S_t$
exactly becomes the $S_h$ term defined above.
In fact, since we are in the broken symmetry phase we can expand
$U_{\vec x;0}$ as follows (since $N_t=1$ we eliminate any reference to
the time coordinate)

\eq
U_{\vec x;0}\equiv e^{i\frac{\phi(\vec x)}{\sqrt{\beta_t}}}
={\bf1}+i\frac{\phi(\vec x)}{\sqrt{\beta_t}}
-\frac{\phi^2(\vec x)}{2\beta_t}
+\cdots
\label{svil}
\en
where $\phi(x)$ is an Hermitian $N \times N$ matrix.
By inserting (\ref{svil}) in $S_t$
 we find:

\eq
S_t(m_0)=\frac{\beta_t}{N}~~tr\left\{d~\Omega_{x}
{\bf 1}+\frac{1}{ \beta_t}\sum_{\vec x} \left(-m_0^2 \phi(\vec x)^2+
\sum_{i=1}^d U_{\vec x;i}\phi(\vec x)U^\dagger_{\vec x;i}
\phi(\vec x+\hat i)\right)\right\}
\label{htkm}
\en
where $m_0^2=d$, $\Omega_x$ is the volume of the
$d$ dimensional space, and we used the periodic boundary conditions and
the fact that $N_t=1$ to identify the two space-like $U$ matrices.

Apart from an irrelevant constant, eq.(\ref{htkm}) coincides with
$S_h(m_0)$ defined in the previous section. The coupling constant
$\beta_h$ has been conventionally fixed to $1$ in eq.(\ref{htkm}).
It depends on the mean value of the Polyakov loops around its
minimum (see~\cite{CAP} for the details), but its precise value turns
out to be irrelevant, the only important quantity being the mass term
$m_0$.

There are some obvious cautions which one must
have in rescaling the $(\beta_t,\beta_s,N_t)$ parameters:
\begin{description}

\item{i]} One must be in the scaling regime (to be defined more
precisely later) to extract the same physics from different lattice
regularizations.

\item{ii]} The rescaling of $N_t$ implies a change of the ratio
$\epsilon$ between time-like and space-like lattice spacings.
In $d=3$, $g$ is adimensional, and such a rescaling requires
a proper $\Lambda(\epsilon)/\Lambda$ correction.

\item{iii]} When $N_t$ is pushed up to the extreme value $N_t=1$ spurious
effects due to lattice artifacts must be expected. Actually, this is
nothing but the lattice counterpart of the decoupling of non-static
modes and reminds us that the naive limit $N_t=1$
corresponds to the  complete dimensional reduction. The problem is
cured
by adding to the  action the $S_{ns}$ term thus obtaining the
correct dimensional reduction.

\end{description}

We still have to fix the region in the $(g^2,T)$ plane where we expect
 our assumptions to hold.
This is not difficult for $g^2$, which we must
require  to be in the region where scaling already
 holds in the zero
temperature regime of the theory. On the contrary we have no general
argument to set a threshold in $T$ which can only be estimated a
posteriori. However it is by now widely accepted that in (3+1) non
abelian L.G.T., dimensional reduction holds for $T>2T_c$. Teper's
results~\cite{t1,t2} suggest that in the (2+1)
dimensional case this threshold can be
set already at $T>T_c$. This fact will play a major role in the comments
of the last section.
Let us stress that the most important consequence of our dimensional
reduction scheme is that the space-time coupling constant, namely the
coupling constant of the dimensionally reduced theory, is {\sl
completely fixed to be} $\beta_s=( 2Na^{d-4})/(g^2T)$, where
$g^2$ is the coupling of the original $(d+1)$ theory. This turns out to
be a very stringent test of the whole approach, and it is the ultimate
reason of its predictive power.

The second step is now to show that the contributions due to $S_h$ and $
S_{ns}$ are negligible if one studies space-like observables. More
precisely, we will show  that their contributions are of order $
1/\beta^2$ and are negligible in the continuum limit $\beta \to \infty$
We shall obtain this result in two steps: first we take into account
only  $S_h$, in which the field $\phi$ appears only
quadratically.
This allows us to integrate the field $\phi$ exactly and
gives the so called ``induced action''  for the space-like gauge
fields~\cite{KM,CAP},

\eq
\int DU D\Phi exp (-S_t(m)) \sim \int DU exp ( -S_{ind}[U])
\en
with:
\eq
S_{ind}[U] = - \frac{1}{2} \sum_{\Gamma} \frac{ |{\rm Tr} U[\Gamma]|^{2}}
{l[\Gamma] (2m_0^2)^{l[\Gamma]}}~~~,
\label{gauge}
\en
where $l[\Gamma]$ is the length of the loop $\Gamma$, $U[\Gamma]$ is
the ordered product of link matrices along $\Gamma$ and the summation
is over all closed loops. It is possible to restrict this summation
to the non-backtracking loops only,
through a suitable renormalization of the
mass term $m_0 \to m_R$. This can be done by ``dressing'' any non-
backtracking closed loop with all possible backtracking paths starting
from a generic site of the loop.

This induced action is very interesting and has been extensively
 studied in the context of the so called Kazakov-Migdal model~\cite{KM}.
In particular it has been shown~\cite{KMSW} that, if no other gauge
self-interaction term is present, near the continuum limit all the loops
equally contribute, one is not allowed to truncate the sum to the
smallest loops, and the resulting theory in the continuum is very
different from ordinary QCD. Moreover, as in the
induced action all gauge fields are in the adjoint representation,
the expectation value of a space-like Wilson loop is exactly zero.
The vanishing of spatial Wilson loops in a pure induced action
indicates that in our case such quantities are dominated by the
space-like part of the Wilson action $S_s$ and that the induced
action can be treated as a perturbation
We shall see in the next section by an explicit calculation in the $d=2$
case, that all contributions of order $1/\beta_s$ coming from $S_{ind}$
cancel exactly and one is left with subleading
 $1/\beta_s^2$ contributions only. The
reason of this result can be easily understood if one
truncates  $S_{ind}$ to its
first  term (we shall see in the next section that in $d=2$ this
truncation already gives a very good approximation of the whole
series). This reduction  is particularly
interesting because the resulting action is exactly of the so called
``fundamental-adjoint'' type. For instance in the $SU(2)$ case:
\eq
S_{f-a}=\sum_{n,i<j}\left\{\frac{\beta_f}{2}{\rm Tr_f}(U_{n;ij})
+\beta_a{\rm Tr_a}(U_{n;ij})\right\},
\en
where  ${\rm Tr_a}$ is
the trace in the  adjoint
representations\footnote{notice the unusual normalization of the adjoint
term, which has
been chosen for future convenience.}.
In our case $\beta_f=\beta_s$ and $\beta_a=1/(2m_R^2)^4$.
 This class of actions was carefully
studied some years ago. It was shown that in $2$ and $3$ dimensions,
near the continuum limit and
for  small values of $\beta_a$ the correction with respect to the usual
Wilson action is completely encoded by a suitable rescaling of
$\beta_f$, for instance in the $SU(2)$ case:
 $\beta_f \to (
\beta_f~ +~ 8\beta_a)$.
Since in the dimensionally reduced theory the coupling
constant is dimensional, any dimensional quantity, say ``Q'',
must scale as $Q=c/\beta_s$ and the above
rescaling implies $Q\sim c/(\beta_s+8\beta_a)
\sim c/\beta_s-8c\beta_a/\beta_s^2$.

Finally, let us consider the effect of $S_{ns}$.
 The quartic term in $S_{ns}$ makes it impossible to integrate
explicitly over the fields $\phi(x)$, however the properties of the
induced action can equivalently be studied by integrating over the gauge
fields  $U_{{\vec x};i}$ in $S_t$.
This can be done by using the well known Itzykson-Zuber
formula~\cite{hcizm}:

\eq
I(\phi(x),\phi(y)) = \int D U \exp \left(
N \, tr \phi(x) U \phi(y) U^{\dagger}
\right) \propto \frac{\det_{ij} \exp(N \lambda_i(x)
\lambda_j(y) )}{\Delta(\lambda(x))
\Delta(\lambda(y))}
\label{izhc}
\en
where $(x,y)$ are nearest neighbour links
of the lattice,
$\lambda_i(x)$ are the eigenvalues of the matrix $\phi(x)$ and
\eq
\Delta(\lambda) = \prod_{i<j} (\lambda_i-\lambda_j)
\label{vandermond}
\en
is the Vandermonde determinant.

Notice that in the SU(2) case,  which is the one relevant in the
following section,
there is only one field $\lambda(x)$. It has been shown that
in this case, and for the range of parameter in which we are interested,
the theory is very well described by a simple mean field
approximation~\cite{ags}. Moreover it has been shown
in~\cite{KMSW,dksw}, how
to deal with the terms coming from the perturbative expansion of $S_s$,
which give origin to the so called ``filled Wilson loops''~\cite{dksw}.
Their contribution can be precisely estimated within the mean-field
approximation.  The main outcome
of this alternative approach is that it allows
to study more general than quadratic potentials and, in
particular, to take into account the contribution of $S_{ns}$.
The important point is that (within the mean field approximation)
all the results obtained above by
integrating first on the $\phi$ field, and also all the steps of the
strong coupling expansion that we shall describe in the next section
remain unaltered. The only effect of the $S_{ns}$ term is to change
the expansion parameter $m_R$, whose value can be evaluated
within the mean field approximation as a function of $h$ and $k$.

\section{SU(2) in (2+1) dimensions}

The case $d=2$ is particularly interesting because in the dimensionally
reduced theory one can use strong coupling expansions to test our
assumption and the various approximations. Moreover, very precise and
careful estimates of all physical quantities relevant to our analysis
have been recently published in the case of the SU(2)
model~\cite{t1,t2}. Let us summarize these results (see~\cite{t1,t2}
for details). In the following $\beta_s=\beta_t=\beta=4/(a~g^2)$ and all
the stated results are intended to be valid near the continuum limit,
namely for large values of $\beta$.
\begin{description}
\item{i]} The zero temperature string tension $\sigma(0)$ behaves as
follows:
\eq
  a\sqrt{\sigma(0)}=\frac{1.336(10)}{\beta}+\frac{1.122}{\beta^2}
\label{sig}
\en
\item{ii]} The deconfinement temperature is:
\eq
    \frac{T_c}{\sqrt{\sigma(0)}}=1.121(8)
\label{tc}
\en
\item{iii]} In the deconfined phase $T>T_c$ the space-like string tension
rises linearly with the temperature according to the law:
\eq
  \sigma(aT)=l_0~\sigma(0)~T
\label{linear}
\en
where $l_0$ is a new physical length (related to the thickness of the
chromoelectric flux tube at zero temperature, see~\cite{t1,cfggv})
 whose value is:
\eq
l_0=\frac{1.22(4)}{T_c}=\frac{1.09(5)}{\sqrt{\sigma(0)}}~~~.
\label{l0}
\en
It is interesting to see that the behaviour of eq.(\ref{linear})
is present in the whole high temperature phase, starting from the
critical point. To be precise, eq.(\ref{linear}) has been tested at
$\beta=9$ (where the critical temperature is between $N_t=5$ and
$N_t=6$) in the whole range $N_t=2-6$.

\end{description}

Inserting eq. (\ref{l0})  in (\ref{linear})
we obtain:
\eq
a~\sqrt{\sigma(0)}=0.92(5) \frac{a~\sigma(aT)}{T}~~~~.
\label{ss}
\en

Let us now assume dimensional reduction, and rescale the above lattices
to $N_t=1$ as described in the previous section. The resulting,
rescaled, and now asymmetric couplings are : $\beta_s=\frac{\beta}{aT}$,
$\beta_t=\beta ~aT$.

Let us first neglect the contribution coming, in the notations of the
previous section,  from $S_{h}$ and $S_{ns}$ and take only into
account the two-dimensional theory given by $S_{s}$. This is exactly
solvable, and the string tension is (see for instance~\cite{dz}):
\eq
a^2~\sigma(aT)=-\log\left(\frac{I_2(\beta_s)}{I_1(\beta_s)}\right)
\en
where $I_n(\beta)$ is the $n^{th}$ modified Bessel function.

By using the large $\beta$ expansion of the Bessel function
we obtain:
\eq
a^2~\sigma(aT)=\frac{3~aT}{2\beta}+\cdots
\label{bessel}
\en
and by inserting this into eq.(\ref{ss}) we find:
\eq
  a\sqrt{\sigma(0)}=\frac{1.38(6)}{\beta}
\label{result}
\en
 which is in remarkable agreement with the known value given in
 eq.(\ref{sig}).
Notice  that within our framework the linear rise of the space-like
string tension with the
temperature also has a natural explanation, since it simply encodes the
$T$-dependence of $\beta_s$ with respect to $\beta$.
The impressive agreement
between (\ref{sig}) and (\ref{result}) is due to the fact that  the
corrections coming from $S_{h}$ and  $S_{ns}$ are of
order $1/\beta^2$ and do not affect the behaviour of eq.(\ref{result}).

The fact that in $d=2$ the strong coupling expansion converges up to the
continuum limit allows us to give some more evidence of this statement,
and more generally  to show the validity of
the approximations made in the last part of the previous section.
In particular we shall show that  $S_h$ and $S_{ns}$ give only
corrections of order $1/\beta_s^2$ to the string tension and moreover
that the coefficient of this $1/\beta_s^2$ term is very small and
actually negligible with respect to other $1/\beta_s^2$ terms of the
expansion.
In order to show this
let us first integrate over the field $\phi$ and let us study the induced
action $S_{ind}$.
Let us select one particular loop $\Gamma$ in $S_{ind}$. By using the
fact
that for SU(2) $|{\rm Tr_f} U[\Gamma]|^{2}={\rm Tr_a} U[\Gamma]+1$
 and by taking into account that the same loop
appears $2~l[\Gamma]$ times (two possible orientations and $l[\Gamma]$
starting points), the corresponding Boltzmann factor can be written as
$~~exp\{\lambda(\Gamma){\rm Tr_a} (U[\Gamma])\}~~$
with $~~\lambda(\Gamma)=1/(2m_R^2)^{\l(\Gamma)}$.
Its character expansion is (setting $\lambda(\Gamma)=\lambda$ for
brevity):
\eq
e^{\lambda~\chi_1(U[\Gamma])}=
e^{\lambda}~[I_0(2\lambda)-I_1(2\lambda)]~[1+\sum_{j=1}^{\infty}
\mu_j(\lambda)\chi_j(U[\Gamma])]
\en
where $\chi_j$ is the character of the $j^{th}$ irreducible
representation
and
\footnote{Notice the different normalization with respect to the usual
character expansions.}
\eq
\mu_j(\lambda)=\frac{I_j(2\lambda)-I_{j+1}(2\lambda)}
{I_0(2\lambda)-I_1(2\lambda)}
\en
so that
\eq
\mu_j(\lambda)~\sim~ \lambda^j~ \sim~(2 m_R^2)^{-j~l[\Gamma]}~.
\label{small}
\en

For any loop $\Gamma$ (of length $l[\Gamma]$ and area $A$)
and for any representation $j$
  the first contribution is given by the
insertion of the loop $\Gamma$ {\sl inside the Wilson loop}
(we assume to be in the limit of infinite area Wilson loops and
neglect boundary corrections). One must take into account the two
possible choices  ($j-1/2$ and $j+1/2$)
for the representation of the space-like plaquettes
inside the loop $\Gamma$ and then subtract the ``excluded
vacuum'' contribution.
The result is:
$$
[\mu_j(\lambda)]\left\{(j+1)\left[\frac{I_{2j+2}(\beta_s)}
{I_{2}(\beta_s)}\right]^A+j\left[\frac{I_{2j}(\beta_s)}
{I_{2}(\beta_s)}\right]^A-(2j+1)\left[\frac{I_{2j+1}(\beta_s)}
{I_{1}(\beta_s)}\right]^A\right\}$$

\eq
{}~\sim~[\mu_j(\lambda)]
\frac{2A^2j(j+1)(2j+1)}{\beta_s^2}
\label{b2a}
\en

As anticipated, both the constant and the $1/\beta_s$ terms vanish in
eq.(\ref{b2a}).

Diagrams in which several different loops $\Gamma$ are simultaneously
present and overlap do not allow a similar compact expression, but also
for them it can be shown, through an iterative elimination of all
possible subdiagrams, that only $1/\beta_s^2$ terms survive.
Let us call $B_{ind}(m^2_R)/\beta_s^2$ the overall contribution (
eq.(\ref{b2a}) plus multiple insertions of $\Gamma$ loops) of
$S_{ind}$ to the order $1/\beta_s^2$ to the string tension.
Even if we cannot give the explicit value of $B_{ind}(m_R^2)$, we can at
least evaluate its order of magnitude.

A first, simple calculation can be made within the strong coupling
expansion:
we have checked by evaluating the first few orders that for $m_R>1.5$
$B_{ind}$ is almost saturated by the terms of eq.(\ref{b2a}). Even more,
it can be seen that the sum is actually dominated by its first term, the
single plaquette contribution.
If we  keep in the expansion only the simple plaquette
($l[\Gamma]=4$, $A=1$), and
restrict ourselves to the $j=1$ representation the contribution is $-12
\mu_1(\lambda)/\beta_s^2\sim -12
/(2m_R^2)^4\beta_s^2$. This is exactly what one would find shifting
$\beta_s \to (\beta_s+8/(2m_R^2)^4)$ in eq.(\ref{bessel}) as described in
the last part of the previous section.

In order to test the order of magnitude of $B_{ind}(m_R^2)$  beyond the
first orders of the strong coupling expansion, let us look at its
contribution to the space-like string tension.
Taking also into
account the next to leading order expansion of the Bessel functions, we
have:

\eq
a^2~\sigma=\frac{3}{2\beta_s}~+~\frac{3}{4\beta_s^2}~
+~\frac{B_{ind}(m^2_R)}{\beta_s^2}~
\label{b2}
\en

It is interesting to notice that the contribution coming from
$S_{ind}$ is opposite in sign to that due to the next to
leading order expansion of the Bessel functions. This gives us
a direct way (by simply
 looking to the sign of a possible $T^2$ term in $\sigma(aT)$)
 to test the order of magnitude of $B_{ind}(m_R^2)$ with respect to
the next  to leading Bessel
correction.

A $T^2$ dependence of $\sigma(aT)$ implies a linear $T$ dependence
in $l_0$. A good parameter to test this behavior is
$\delta l_0 (aT)= l_0(aT)/l_0(aT_c)$, which we can estimate to be:
\eq
\delta_{th} l_0 (aT)= \left( 1+\frac{a(T-T_c)}{2\beta}
-\frac{2B_{ind}(m^2_R)a(T-T_c)}{3\beta}\right)
\en
where the second term comes from the next to leading order expansion of
the Bessel function and the last term is
contribution of $S_{ind}$.

This behaviour can be tested with a set of very precise data on
$l_0(T)$ obtained by M. Teper at $\beta=9$~\cite{pc}. The results are
collected in
Tab.1; $\delta_{exp}l_0$ is normalized at  $aT=1/6$~. In order
to make contact with eq.(\ref{l0}), one must use the fact that at $\beta
=9$, $1/(aT_c)=5.65$~\cite{t1,t2}.

%     TAB 1
\vskip0.3cm
$$\vbox {\offinterlineskip
\halign  { \strut#& \vrule# \tabskip=.5cm plus1cm
& \hfil#\hfil
& \vrule# & \hfil# \hfil
& \vrule# & \hfil# \hfil &\vrule# \tabskip=0pt \cr \noalign {\hrule}
&& $N_t$ && $l_0/a$   && $\delta_{exp} l_0$   &
\cr \noalign {\hrule}
&& $2$ && $7.21(12)$ && $1.062(30)$ & \cr \noalign {\hrule}
&& $3$ && $7.32(7)$ && $1.078(22)$ & \cr \noalign {\hrule}
&& $4$ && $7.08(6)$ && $1.043(20)$  & \cr \noalign {\hrule}
&& $5$ && $6.92(8)$ && $1.019(20)$  & \cr \noalign {\hrule}
&& $6$ && $6.79(8)$  && $1$ & \cr \noalign {\hrule}
}}$$
\begin{center}
{\bf Tab.I.}{\it~~ $l_0$ and $\delta l_0$ as  a function of the inverse
temperature $N_t=1/aT$ in the (2+1) dimensional SU(2) LGT at $\beta=9$.}
\end {center}
\vskip0.3cm

 It is possible to see that the $T$-dependence of $l_0$ has exactly
the same sign and order of magnitude of the next to leading term of
the Bessel function and, as a consequence of this,
that $B_{ind}(m^2_R)$ must be very small and that it
 is well approximated by its first term in the strong coupling
expansion even if all orders are taken into account.

\section{SU(2) in (3+1) dimensions}

In $d=3$ the validity of dimensional reduction has been already well
established by G. Bali and collaborators in~\cite{bfhks}.
By using the set of very precise determinations of $\sigma(aT)$ at $\beta=
2.74$ shown in Tab.II,

%     TAB 2
\vskip0.3cm
$$\vbox {\offinterlineskip
\halign  { \strut#& \vrule# \tabskip=.5cm plus1cm
& \hfil#\hfil
& \vrule# & \hfil# \hfil &\vrule# \tabskip=0pt \cr \noalign {\hrule}
&& $T/T_c$ && $\sqrt{\sigma}/T_c$  &
\cr \noalign {\hrule}
&& $2$ && $1.97(3)$  & \cr \noalign {\hrule}
&& $2.67$ && $2.43(3)$   & \cr \noalign {\hrule}
&& $4$ && $3.28(2)$   & \cr \noalign {\hrule}
&& $8$ && $5.70(4)$   & \cr \noalign {\hrule}
}}$$
\begin{center}
{\bf Tab.II.}{\it ~~Space-like string tension as a function of the
temperature in the (3+1) dimensional SU(2) LGT at $\beta=2.74$, taken
from ref.~\cite{bfhks}.}
\end {center}
\vskip0.3cm

they were able to show
 that the space-like string tension behaves, for $T>2T_c$
 as
\eq
\sqrt{\sigma(aT)}=(0.334\pm 0.014)~g^2(T)T
\label{d=3}
\en
with
\eq
g^{-2}(T)=\frac{11}{12\pi^2}log(T/\Lambda_T)
\label{g}
\en
with $\Lambda_T=0.050(10)T_c$
\footnote{We have chosen the one loop formula of ref.~\cite{bfhks}
 so as to compare with the $\Lambda(\epsilon)/\Lambda$
correction (see next section) which is also a one loop result}.
This result is in remarkable agreement
with eq.(\ref{sig}) (remember that $\beta=4/g^2$). This is indeed
an impressive test of dimensional reduction. The only remaining problem
is a discrepancy of a factor of two between the coupling constant $g^2(T
)$, as extracted from (\ref{g}), and the value quoted in ref.~\cite{lmr}.

The only improvement
that our analysis can add to this result is  to clarify this last
issue. The point is that,
 as we remarked earlier, P.Lacock and
collaborators  define their
reduced theory exactly as if they had chosen $N_t=1$. This means
that if we want
to compare the coupling constant $g^2(T)$ with that of
ref.\cite{lmr},  we must estrapolate $N_t \to 1$. This can be achieved
by using eq.s(\ref{d=3},\ref{g}).  One obtains
$\sqrt{\sigma(aT=1)}/T_c=9.97(8)$ which would correspond to $g^2=1.86$.
This result is still rather far from the value\footnote{Obtained by
interpolating the
two results at $\beta_4=2.80$ and $\beta_4=2.50$ (in their notations)
quoted in~\cite{lmr}).}: $g^2\sim 1.31$
 quoted by P.Lacock and
collaborators.
But this is not the end of the story. If we want to make contact with
the dimensionally reduced theory we must compare our data with
the  {\sl exact} behaviour of the string tension of the three
dimensional SU(2) model, namely we must also take into account the $1/
\beta^2$ term in eq.(\ref{sig}), which, due to the fact that $\beta$
is not very large, turns out to be rather important. This is similar to
what we did in the $d=2$ case taking into account the next to leading
order of the Bessel function.
Inserting our value
of $\sqrt{\sigma}$ in eq.(\ref{sig}), and solving with respect to
$\beta$ we find $\beta=2.79(6)$, namely $g^2=1.43(3)$, where the quoted
error takes also into account the uncertainty in the second coefficient
of eq.(\ref{sig}). The main reason
of interest of this result is that it is a further, independent,
cross-check of the proposed dimensional reduction framework, since it
coincides, within the errors, with the value of the coupling in
the (3+1) dimensional model. Moreover it almost
fills the gap with the value of $g^2$ found by P.Lacock and collaborators.

As a last remark let us notice that the corrections coming from $S_{h}$ and
$S_{ns}$ are again very small and essentially negligible.
 Let us again truncate
 the induced action to its first term, then its contribution can
be encoded in the shift  $\beta \to \beta+8/(2m^2_R)^4$.
In this case, thanks to~\cite{lmr}, precise
values of the constants $h$ and $k$ are
known, and $m^2_R$ turns out to be $m^2_R\sim 2.3$ .
Inserting this value it is easy to see that the
the correction is  smaller than the quoted error of
$\beta$.

\section{Conclusions and Speculations.}

In this paper we have shown that a simple dimensional reduction
scheme can describe rather well space-like observables at high
temperature. We gave arguments to support the conjecture that only the
space-like part of the original action (after a suitable rescaling of
the couplings) is relevant for the result.
All the results quoted in the previous sections were rather
straightforward applications of well known ideas (dimensional reduction)
and techniques (Migdal-Kazakov approach to induced actions).

Let us now conclude with two more speculative, although very suggestive
considerations.

\vskip 1cm
{\sl Determination of $\Lambda_T$}
\vskip 0.2cm

The crucial difference between the $d=2$ and $d=3$ cases is that in
 $d=3$, $g$ is adimensional, and the rescaling of $N_t$ requires
a proper $\Lambda(\epsilon)/\Lambda$ correction.
While in $d=2$ we had that $\sqrt{\sigma(aT)}/T$ was constant, namely
 $\frac{\sqrt{\sigma(aT_1)}}{T_1}\frac{T_2}{\sqrt{\sigma(aT_2)}}=1$,
we now expect (assuming $T_1>T_2$) :
\eq
R\left(\frac{T_1}{T_c},\frac{T_2}{T_c}\right)\equiv
\frac{\sqrt{\sigma(aT_1)}}{T_1}\frac{T_2}{\sqrt{\sigma(aT_2)}}=
\Lambda(\frac{T_1}{T_2})/\Lambda~~.
\label{ratio}
\en
 The $\Lambda(\epsilon)/\Lambda$ correction was
studied by F. Karsch in~\cite{k}. He was able to obtain in the $(g\to
0,\epsilon$ fixed) limit the following result

\eq
log\left(\frac{\Lambda(\epsilon)}{\Lambda}\right)=
-\frac{c_\sigma(\epsilon)+c_\tau(\epsilon)}{4b_0}~~~,
\label{k}
\en
where $c_\sigma(\epsilon)$ and $c_\tau(\epsilon)$ are two known
functions
(see~\cite{k}) whose main property is that in the limit $g\to 0$:
\eq
\frac{\partial c_\sigma}{\partial \epsilon}\vert_{\epsilon=1}+
\frac{\partial c_\tau}{\partial \epsilon}\vert_{\epsilon=1}=b_0
\label{limit}
\en
with $b_0=11/24\pi^2$.
Some particular values of eq.(\ref{k}) are:
$~~~\Lambda(4/3)/\Lambda\sim 0.925$,
$~~~\Lambda(3/2)/\Lambda\sim 0.90$,
$~~~\Lambda(2)/\Lambda\sim 0.84$~~.

By using the values of $\sigma(T)$ listed above,
we can construct the following ratios: $R(4,2)=0.832(17)$,
 $R(8,4)=0.869(12)$  $R(2.67,2)=0.924(20)$  $R(4,2.67)=0.901(15)$
 which are in good agreement with the
corresponding values of $\Lambda(\epsilon)/\Lambda$.

We want to stress however that this agreement should be taken with some
caution.
 The main point is that the two scales
$\Lambda_T$ and $\Lambda(\epsilon)$ have a completely different origin.
While eq.(\ref{g}) is a perturbative result in $g^2$, and as
far as $T$ is much larger than $\Lambda_T$ it correctly describes the
$T$ dependence of $g$ and consequently of $\sigma$ for any $T$; our
result is perturbative both in $g^2_\sigma$ and $g^2_\tau$ (in the
notations of ref.~\cite{k}), namely in the space-like and time-like
couplings. This means that in our case $T$ must be as small as possible
if we want to correctly describe the data.

This can be very explicitly seen if we try to extract from the known
values of $\Lambda(\epsilon)/\Lambda$ the scale $\Lambda_T$.

This can be done by equating
\eq
\Lambda(\frac{T_1}{T_2})/\Lambda
=log\frac{T_2}{\Lambda_T}/log\frac{T_1}{\Lambda_T}
\label{ratio2}
\en

The values of $\Lambda_T$ obtained in this way depend on $T_1$ and
$T_2$. We obtain: $\Lambda_T(2,2.67)=0.057 T_c$,~~~
 $\Lambda_T(2,4)=0.053 T_c$~~~
 $\Lambda_T(2.67,4)=0.070 T_c$~~~
 $\Lambda_T(4,8)=0.105 T_c$~.

As expected these values of $\Lambda_T$
 are in good agreement with the one
$\Lambda_T=0.050(10)T_c$ found by G. Bali and collaborators
if small temperatures are chosen, but they rapidly grow as larger
$T$'s are chosen.
Notice, as a side remark, that the two loop value of $\Lambda_T$
reported in ref.~\cite{bfhks} is $\Lambda_T=0.076(13)T_c$, thus
indicating that as higher perturbative terms are taken into account,
the agreement between the two approaches can be extended at higher
temperatures.

Let us conclude by noticing that the above result, namely the possibility
to predict the value of the scale $\Lambda_T$ from the known value of
$\Lambda(\epsilon)/\Lambda$ is quite general, it can be extended to any
gauge group SU(N) and can be written in a rather elegant form in the
$g\to 0$ limit.
 Making an
expansion for small rescaling $\epsilon=T_1/T_2\sim 1$ and
choosing a reference temperature $T_{ref}$, which should be the
 smallest possible value of $T$  compatible with the various
thresholds of the problem, we can use eq.(\ref{limit}) and
we obtain $\Lambda_T\sim  e^{-4}T_{ref}$. Assuming, as reference
temperature the threshold $T_{ref}=2T_c$ at which the scaling behaviour
of eq.(\ref{g}) is expected to hold, we find:
$\Lambda_T\sim 2 e^{-4}T_c~ =~ 0.37~T_c$

\vskip 1cm
{\sl $T_c$ and $\sqrt{\sigma(0)}$ in three dimensions}
\vskip 0.2cm

In the (2+1)  case dimensional reduction
predicts that the space-like string tension must grow linearly with T.
It is rather interesting to notice that, in this case, this behaviour
sets in already at $T=T_c$, thus suggesting that the whole deconfined
phase could be described using dimensional reduction.
This leads to the following speculation.
Let us assume the idealized picture in which the space-like string
tension is exactly constant as a function of $T$ in the region $T<T_c$,
and then at $T=T_c$ sharply starts to rise linearly\footnote{
It is easy to see looking at Teper's data that this change of behaviour
is not so sharp, notwithstanding this,  the idealized picture described
above is only 20\% away from the real behaviour.}.
This implies the relation:
\eq
\sigma(0)=aT_c\sigma(aT=1)=\frac{3T_c}{2a\beta}
\en
which can be easily generalized to the case of SU(N):
\eq
\sigma(0)=\frac{(N^2-1)T_c}{2a\beta}
\label{r1}
\en

Let us now assume the following adimensional relation~\cite{ol}
\eq
\sigma(0)=\frac{\pi}{3}T_c^2
\label{r2}
\en
which comes
from the completely different context
of the effective string approach to the interquark potential in LGT's.
This relation does not depend of the gauge group but only on the number
of space-time dimensions of the model, hence it holds unchanged for all
$N's$. In the SU(2) case it is
 again verified only within 15\% (see the comment at the
end of~\cite{t2} on this point).
Combining (\ref{r1}) and (\ref{r2}) we can predict the value of the
(zero temperature) string tension and the critical temperature for SU(N)
gauge theories in (2+1) dimension as a function of the coupling constant
$\beta$.
\eq
\sigma(0)=\frac{3~(N^2-1)^2}{4\pi\beta^2}
\en
\eq
T_c=\frac{3~(N^2-1)}{2\pi\beta}
\label{tc2}
\en

Up to our knowledge, there is  only one numerical result
(besides those on SU(2)) against which we can  test these conjectures.
It  is the critical coupling at which the SU(3)
model with $N_t=2$ undergoes the deconfinement transition, which turns
out to be $\beta_c\sim 8.17$~\cite{ctdw}.
Setting $N=3$ and $T_c=2$ in eq.(\ref{tc2}) we
find $\beta=7.64$ which is only 10\% away from the numerical result.

It would be quite interesting to have some further independent checks
of these predictions.

\vskip 1cm
{\bf 1. Acknowledgements}

We thank F.Gliozzi, F. Karsch, L.Magnea, S.Panzeri, M. Teper,
and S.Vinti for many helpful discussions. In particular we thank
M. Teper for sending us his results on the SU(2) string tension.

\end{document}